\documentclass[12pt,amsmath,amssymb]{article}
\usepackage{amssymb}
\def\beq{\begin{equation}}
\def\eeq{\end{equation}}
\def\d{\partial}
\def\l{\left(}
\def\r{\right)}

\begin{document}
\begin{center}
  {\Large\bf Brane-induced gravity }\\
  {\Large\bf in more than one extra dimensions:}\\
  {\Large\bf violation of equivalence principle and ghost.} \\
  \medskip S.~L.~Dubovsky and V.~A.~Rubakov\\
    \medskip
  {\small
     Institute for Nuclear Research of
         the Russian Academy of Sciences,\\  60th October Anniversary
  Prospect, 7a, 117312 Moscow, Russia\\
  }
\end{center}
\begin{abstract}

We consider brane-induced gravity model in more than one extra
dimensions, regularized by assuming that the bulk gravity is soft in
ultraviolet. We study linear theory
about flat multi-dimensional space-time
and flat brane. We first find that this model allows for 
violation of equivalence between gravitational and inertial masses of
brane matter. We then observe that the model has a scalar ghost
field localized near the brane, as well as quasi-localized massive graviton.
Pure tensor structure of four-dimensional gravity on the brane at
intermediate distances is due to the cancellation between the extra
polarization of the massive graviton, and the ghost. This is
completely analogous to the situation in the GRS model.

\end{abstract}

\section{Introduction and summary}
 
In view of the observation \cite{Randall:1999vf}
that gravity may be localized on a brane embedded in 
space with one extra dimension of infinite size,
it is of interest to study whether there exist mechanisms
of (quasi-)localization of gravity in spaces with more than one
infinite extra dimensions. 
One proposal of this sort has been put forward in 
Refs.~\cite{DG}. The basic idea~\cite{DGP} is that radiative effects
due to matter residing on the brane  may induce new terms in the
effective action of multi-dimensional gravity 
(cf. Ref.~\cite{sakharov}),
which concentrate on the brane and dominate the gravitational
interactions of brane matter. Thus, the effective action has the form
\beq
\label{*}
S_{tot} = S_{bulk}^{eff} + S_{brane}
\eeq
Here the bulk term involves $(4+N)$-dimensional metric $g_{AB}$
($N>1$
is the numer of extra dimensions)  and at low energy reduces to the
$(4+N)$-dimensional Einstein--Hilbert action\footnote{Leaving aside
the issue of the cosmological constant.} with the fundamental scale
$M_*$. The brane Einsten-Hilbert 
term, on the other hand, involves induced
$4$-dimensional metric $g_{\mu \nu}$ on the brane
and has its own mass scale $M_{Pl}$, which supposedly is 
determined by
dynamics on the brane. It has been argued in Ref.~\cite{Dvali:2001gx} 
that the two scales may be completely different, and, in particular,
that the relation
\beq
\label{**}
   M_* \ll M_{Pl}
\eeq
may hold.  

For more than one extra dimensions, $N>1$, the model exhibits a
potentially interesting UV-IR mixing. Naively, one would expect that
at large distances along the  brane, the relevant terms in $S_{bulk}$
and $S_{brane}$ are the multi-dimensional and four-dimensional
Einstein-Hilbert terms respectively, while the brane may be treated as
$\delta$-functional in transverse directions. This is not the case,
however, because of the singularity of the $N$-dimensional propagator
\cite{Dubovsky:2001pe,Kiritsis:2001bc,1}. Hence, the behavior of the
model at large distances along the brane depends on how the
singularity in transverse dimensions is resolved.

One way to resolve this singularity would be to smear the
$\delta$-function in the brane action. This proposal, however, suffers
the strong coupling problem at unacceptably low energies
\cite{Dubovsky:2001pe}. Hence, we will not consider this option any
longer.

Another proposal \cite{Kiritsis:2001bc,1}
which is the subject of this paper\footnote{Just for brevity,
we will call this proposal as ``brane induced gravity'' in what
follows.}, is that the bulk gravity is
``soft'' at distances shorter than $M_*^{-1}$.
Under this assumption, matter on the brane experiences
four-dimensional gravity at intermediate distances~\cite{Kiritsis:2001bc,1}
\beq
   \frac{1}{M_*} \ll r \ll r_c \equiv \frac{M_{Pl}}{M_*^2}
\label{I3*}
\eeq
while four-dimensional Newton's law ceases to hold at both  
short and
ultra-large distances. It is worth noting that 
this multi-dimensional ``brane-induced gravity'' model, 
linearized
about flat background, leads to pure tensor~\cite{DG}
four-dimensional gravity on the brane at intermediate distances
(\ref{I3*}), without an extra scalar inherent in the linearized 
brane-induced gravity in one extra dimension~\cite{DGP}.

These features make brane-induced gravity with $N>1$ potentially
interesting, in particular, from the viewpoint of the 
cosmological constant problem~\cite{coscon}. The violation of
four-dimensional Newton's law at ultra-large distances, combined with
the absence of extra scalar interaction
on the brane at intermediate scales is alarming, however, as the same
property was present in the model of Ref.~\cite{GRS}
which has been found to have a ghost~\cite{Csaki,Pilo:2000et}.
Hence, brane-induced gravity in more than one extra dimensions
is worth studying in some detail.

In this paper we consider brane-induced gravity, 
linearized about flat
multi-dimensional space and flat brane, 
mostly at $N>2$; we discuss somewhat special
case $N=2$ towards the end. In Section 2 we neglect complications due
to tensor structure, and study a scalar counterpart of the model.
We find that once the brane has finite thickness, the equivalence
between gravitational and inertial masses is generally violated for
matter on the brane, even in theory restricted to intermediate scales
(\ref{I3*}). This is again alarming, since in other
models\footnote{Leaving aside models with extra light four-dimensional
degrees of freedom.}, violation of
``charge universality'' (in the gravitational context,
equivalence principle) is a signal
for inconsistency~\cite{Dubovsky:2001pe}. 

We then proceed in  Section 3 to brane-induced gravity itself.
We study the linearized field equations, assuming first that the bulk
term has the tensor structure of General Relativity. We begin with the
study of low-mass states which are localized or quasi-localized near
the brane. We find in Section 3.1 that one such state is a
four-dimensional scalar; it is exactly localized on the brane and has
negative (tachyonic) mass squared. Another state is a massive
four-dimensional graviton\footnote{The graviton has finite, 
though very small width,
$\Gamma_{graviton} \ll m_{graviton}$, i.e., it is, strictly
speaking, quasi-localized.}. Both masses are of order
\[
   |m_{tachyon}| \sim |m_{graviton}|
 \sim r_c^{-1} \equiv \frac{M_*^2}{M_{Pl}}
\]
%The graiton has finite, but very small width,
%$\Gamma_{graviton} \ll m_{graviton}$, i.e., the
%graviton is, strictly
%speaking, quasi-localized. 
In Section 3.2 we proceed to show
that the tachyon is actually a ghost. This can be seen in two
ways. One is to study the propagator of the full linearized theory
near the tachyon pole and show  that the residue has negative sign.
Another way is to evaluate the propagator from brane to brane, which
describes gravitational interaction  of matter on the brane. We find
that the
brane-to-brane propagator is a sum of two terms, one of which has a
pole at $p^2 = m_{graviton}^2$ with tensor structure appropriate for
massive graviton, while another is a scalar ghost term (of overall
negative sign) with a pole at $p^2 = m_{tachyon}^2$. This situation is
completely analogous to that in the model of Ref.~\cite{GRS}: at
intermediate scales (\ref{I3*}), the ghost term cancels out the 
extra~\cite{VDVZ} scalar
part of the massive graviton propagator, so that the brane-to-brane
propagator at intermediate distances has massless tensor form.

We comment on the case of two extra dimensions, $N=2$, in Section
3.3. There are peculiarities, but the outcome is the same: the model has
tachyonic ghost.

In Section 4 we generalize by allowing for most general tensor
structure of the linearized bulk equations (in fact, there are only
two terms consistent with $(4+N)$-dimensional
general covariance).
We again study the case $N>2$, and evaluate the brane-to-brane
propagator. We find that it again has ghost term, although the mass of
the ghost is no longer necessarily tachyonic.

Our overall
conclusion is that the linearized
brane-induced gravity as it stands
has a ghost, if the number of extra dimensions is larger than one.
We  interprete this property as an indication  that this version of
induced gravity cannot
emerge as a low energy limit of any consistent microscopic theory.

It is worth noting that the low energy action of the general form
(\ref{*}) emerges in string theory context
\cite{Antoniadis:2002tr}. Furthemore, the hierarchy (\ref{**}) is also
possible in string theory framework \cite{Antoniadis:2002tr}. It would
be of interest to understand how string theory resolves the UV-IR
ambiguity inherent in the case of more than one extra dimensions.
\section{Scalar model}
%\subsection{Scalar propagator}

We begin with a counterpart of the brane-induced gravity
with metric perturbations  mimicked by a single scalar field 
$\Phi$.
In what follows it will be convenient to consider thick brane, and
take the limit of delta-function brane in the end of calculations, if
desired.
It has been argued in Ref.~\cite{1}
that the 
loops (and/or non-perturbative effects) involving matter
on the brane induce non-local terms in the effective action,
with the scale of non-locality set by the brane thickness
$\Delta$. At quadratic level, this effect is modelled by the
 induced action of the following form~\cite{1},
\beq
    S_{brane}^{(2)} = \frac{M_{Pl}^2}{2} \int~d^4x~d^N y d^N y' 
f^2 (y) \partial_{\mu} \Phi (x,y)
f^2 (y') \partial^{\mu} \Phi (x,y')
\label{induced}
\eeq
where $f(y)$  is a smooth function localized near the brane; it
accounts for thickness of the brane. 
$N$ is the number of extra dimensions; we concentrate on the case
$N>2$.
Without loss of generality,
$f$ is normalized to unity,
\beq
  \int~d^N y ~f^2(y) =1
\label{normf}
\eeq
and is non-zero in a region of size of order
$\Delta$.
Hereafter $X_A = (x_\mu , y_a)$ are coordinates in $(4+N)$~dimensions,
$\mu = 0, \dots, 3$; $a=4,\dots N+3$; signature of metric is mostly 
negative. 

Let the bulk theory
have the effective action $S^{eff}_{bulk} [\Phi]$.
There are two more assumptions in the model~\cite{1}: 
(i) The mass scale entering $S^{eff}_{bulk}$
is $M_*$ which is much smaller than $M_{Pl}$; (ii) The bulk theory is
``soft'' at length scales below $1/M_*$, which we understand as 
the assumption that the Green's functions of the bulk theory
rapidly vanish at high Euclidean momenta.
 
For considering linearized 
theory (weak sources), let us neglect the non-linear terms in the
bulk effective action. Then the only relevant
term in $S^{eff}_{bulk}$ is quadratic in $\Phi$, and has the form
\[
  S^{eff, (2)}_{bulk} = - \frac{1}{2}  
\int~d^{N+4} X  \Phi (X) {\cal F}(\Box^{(4+N)})
\Box^{(4+N)}  \Phi (X)
\]
where ${\cal F}(\Box^{(4+N)}) \Box^{(4+N)}$
 is the exact inverse propagator of the bulk theory.
At low energies, the form factor ${\cal F}$ is a constant
of order $M_*^{2+N}$ (note that the field $\Phi$ is
dimensionless). Let us denote the
exact propagator of the bulk theory by
$D_* (X - X')$, so that
\[
   D_*(P) = \frac{1}{P^2 {\cal F}(-P^2)}
\]
where $P^2 = p^2 - p_y^2$, and $p^2=p_\mu p^\mu$. 
  At momenta below $M_*$, the propagator $D_*$ coincides with
the free propagator,
\beq
  D_* (P) = \frac{1}{M_*^{2+N}} \frac{1}{P^2} \; , \;\;\;
  |P^2| \ll M_*^2
\label{Das}
\eeq
and, by assumption of
softness, $D_*(P)$ rapidly tends to zero at large negative (Euclidean)
$P^2$, with
characteristic scale $M_*$. 

\subsection{Scalar propagator}
Let the source on the brane be characterized by a spread
function $g^2(y)$, and be $\delta$-function in $x$-coordinates, where, again
without loss of generality, $g$ is normalized to unity,
\beq
  \int~d^N y ~g^2(y) =1
\label{normg}
\eeq
It is convenient to work in mixed representation,
momentum in four dimensions  and coordinate in extra dimesnions. 
One has the following equation for 
the propagator from brane to everywhere for given shape of the source,
\beq
  - {\cal F}(\Box^{(4+N)}) \Box^{(4+N)} G_g(p,y')
  + M_{Pl}^2 p^2 f^2(y) \int~d^N y' f^2(y') G_g(p,y') = g^2(y)
\label{6}
\eeq
where $  \Box^{(4+N)} = -p^2 - \partial_y^2$.

Dvali~et.al.~\cite{1} 
proceed under assumption that $g^2(y) = f^2(y)$
with corrections suppressed by $M_*/M_{Pl}$. Let us drop this
assumption, and see what happens.

Equation (\ref{6}) has the following solution,
\beq
    G_g(p,y) = D_g (p,y) - 
\frac{M_{Pl}^2 p^2 D_{fg}(p)}{1+ M_{Pl}^2 p^2 D_{ff}(p)}
\cdot D_f(p,y)
\label{field}
\eeq
where for any function $u(y)$ one defines
\beq
   D_u (p,y) = \int~d^N y' ~ D_*(p,y-y') u^2(y')
\label{Du}
\eeq
and for two functions $u(y)$, $v(y)$ one writes
\beq
  D_{uv} (p) = D_{vu}(p)
=  \int~d^N y d^N y' ~ D_*(p,y-y') u^2(y') v^2 (y)
\label{Duv}
\eeq
Let us rewrite the expression (\ref{field}) in the following suggestive 
form,
\begin{eqnarray}
G_g(p,y) &=& \frac{D_g(p,y)}{1 + M_{Pl}^2 p^2 D_{ff}(p)}
\nonumber \\
  &+& 
\frac{M_{Pl}^2 p^2 [D_{ff}(p) D_g (p,y) - D_{fg}(p) D_f (p,y)]}{
 1 + M_{Pl}^2 p^2 D_{ff}(p)}
\label{basic}
\end{eqnarray}
Now, recall that
\[
  D_* (p, y-y') = \int ~d^N p_y~ D_*(p^2 - p_y^2)
\mbox{e}^{ip_y \cdot (y-y')}
\]
To evaluate the integral of the form $D_{uv}(p)$,
we assume that the brane thickness $\Delta$ is much smaller than
$1/M_*$, and write
 for small $y$ and $y'$
\[
  D_* (p, y-y') = D_*^{(0)}(p) + D_*^{(2)} (p) \cdot (y-y')^2 + \dots
\]
Clearly,
\[
D_*^{(0)}(p) = \int~d^N p_y ~D_* (p^2 - p_y^2)
\]
and
 \beq
D_*^{(2)}(p) = - \frac{1}{2N}\int~d^N p_y ~p_y^2 D_* (p^2 - p_y^2)
\label{D2}
\eeq
We assume that the latter integrals are convergent at negative
(Euclidean) four-momenta, $p^2\leq 0$,
because of softness of the propagator $D_*$ at short distances.
On dimensional grounds
\beq
  D_*^{(0)}(|p| \ll M_*) \sim \frac{1}{M_*^4}
\label{es1}
\eeq
and
\beq
D_*^{(2)}(|p| \ll M_*) \sim \frac{1}{M_*^2}
\label{es2}
\eeq
To the first non-trivial order in brane thickness,
one has (assuming that $u$ and $v$ are normalized to unity,
see eqs.~(\ref{normf}) and (\ref{normg}))
\beq
  D_{uv}(p) = D_*^{(0)} (p) +  D_*^{(2)} (p) \Delta_{uv}^2
\label{interm1}
\eeq
where
\[
  \Delta_{uv}^2 = \int~d^N y d^N y'~(y-y')^2 u^2(y) v^2 (y')
\]
explicitly depends on the shapes of the functions
$u(y)$ and $v(y)$ and 
is generically
of the order of $\Delta^2$.

At low momenta, 
$|p| \ll M_*$, one can set $D_{ff} = \mbox{const} \sim M_*^{-4}$ in the
denominators in eq.~(\ref{basic}). Then at intermediate distances
(\ref{I3*}),
virtuality $p^2$ is large enough, and one has 
\begin{eqnarray}
G_g(p,y) &=& \frac{D_g(p,y)}{M_{Pl}^2 p^2 D_{ff}(p)}
 \nonumber \\
&+& \frac{D_{ff}(p) D_g (p,y) - D_{fg}(p) D_f (p,y)}{D_{ff}(p)}
\label{basic2}
\end{eqnarray} 
This propagator determines the field induced by a weak source of
shape $g^2(y)$ in transverse directions, in the theory restricted
to intermediate scales (\ref{I3*}).

\subsection{Potential on the brane}

The interaction between sources with spread functions $g^2(y)$
and $h^2(y)$ is  described by the effective
four-dimenstional propagator, which is the convolution of $G_g(p,y)$
and $h^2(y)$. At intermediate values of momenta,
$M_*^2 \gg p^2 \gg r_c^{-2}$, one has from eq.~(\ref{basic2})
\beq
  G_{eff,~4d} (p) 
= \frac{D_{gh}(p)}{M_{Pl}^2 p^2 D_{ff}(p)}
+ \frac{D_{gh}(p)D_{ff}(p)
- D_{fg}(p)D_{fh}(p)}{D_{ff}(p)}
\label{all}
\eeq
Keeping terms of order $\Delta^2$, one finds
\begin{eqnarray}
 G_{eff,~4d} (p) &=& \frac{D_*^{(0)} (p) + 
D_*^{(2)}(p) (\Delta_{gh}^2 - \Delta_{ff}^2)}{M_{Pl}^2 p^2 D_*^{(0)} (p)}
\nonumber \\
&+&  
D_*^{(2)}(p)(\Delta_{gh}^2 + \Delta_{ff}^2
- \Delta_{fg}^2 - \Delta_{fh}^2)
\label{all2}
\end{eqnarray}
Consider the first term. Because of explicit $p^2$ in denominator,
one can replace
$D_*^{(0)}$ and $D_*^{(2)}$ by constants
at $p\ll M_*$, i.e.
at distances larger than $M_*^{-1}$.
This leads to four-dimensional
Newton's potential with non-universal gravitational constant
\[
    G_{Newton,~eff} = \frac{1}{M_{Pl}^2}\left[ 1
+ \frac{D_*^{(2)}(0)}{D_*^{(0)}(0)} (\Delta_{gh}^2 -
\Delta_{ff}^2)\right]
\]
According to eqs.~(\ref{es1}) and (\ref{es2}), the non-universal 
correction is of order $\Delta^2 M_*^2$. 

This is the main result of this Section: the model allows for (weak)
violation of the equivalence principle, since the spread functions
$g^2(y)$ and $h^2(y)$ may have different shapes, depending on the type of
matter residing on the brane.

The second term in eq.~(\ref{all2}) corresponds to short-ranged force.
According to eq.~(\ref{D2}), one has, in coordinate representation,
\[
  D_*^{(2)}(x) = \frac{1}{2N} 
%\frac{\partial^2}{\partial y^a \partial y^a} 
\d_y^2D_* (x^2 - y^2)\vert_{y=0}
\]
At relatively large distances, $r \gg M_*^{-1}$, the propagator
$D_*$ is a free propagator in $(4+N)$ dimensions, up to a factor
$1/M_{*}^{2+N}$. This gives
\[
  \int~dx^0 D_*^{(2)}(x) = \frac{1}{M_*^{2+N} |{\bf x}|^{3+N}}
\; , \;\;\; \; |{\bf x}| \gg M_*^{-1}
\]
up to numerical constant of order one. Hence, the correction to Newton's
potential  is 
\[
\Delta V (r) = 
\frac{1}{M_*^{2+N} r^{3+N}}
(\Delta_{gh}^2 + \Delta_{ff}^2
- \Delta_{fg}^2 - \Delta_{fh}^2)
\]
This is a short-ranged potential, ``fifth force'',
which again depends on the composition
of matter (functions $g^2$ and $h^2$). It is worth noting that the
latter non-universality exists also at $N=1$ \cite{Kiritsis:2001bc},
where the brane-induced gravity does not show any inconsistency.

\section{Tachyonic ghost}

%\subsection{Wave function of the localized state: tachyon}

Let us now consider the linearized 
brane-induced gravity and keep track of its tensor structure.
In this Section we assume for simplicity that the tensor 
structure of the linearized equations in the bulk  coincides with 
that in the linearized
Einstein theory in $(4+N)$ dimensions. Then the linearized 
field equation takes the following form,
\beq
   {\cal F}(\Box^{(4+N)}) G_{AB} (x,y) + M_{Pl}^2 f^2(y)
\int~dy'~f^2(y') G^{(4)}_{AB} (x, y') = T_{AB}(x,y)
\label{maineqn}
\eeq
where $G_{AB} = R_{AB} - (1/2)g_{AB} R$ is the linearized 
Einstein tensor in $(4+N)$ dimensions, 
$G^{(4)}_{aB} =0$,
the four-dimensional Einstein tensor
$G^{(4)}_{\mu\nu}$ is constructed in terms of four-dimensional
components of the metric. The form-factor
${\cal F}$ has the same properties as above.
The function $f^2(y)$ is again
the spread function for
induced term.

Let us impose the harmonic gauge
\beq
     \partial_A h_B^A = \frac{1}{2} \partial_B h^{A}_A
\label{gauge}
\eeq
where $h_{AB}$ are perturbations about Minkowski metric $\eta_{AB}$;
indices are raised and lowered by Minkowski metric.
Then one has
\beq
   G_{AB} = - \frac{1}{2} \Box^{(4+N)} (h_{AB} - \frac{1}{2}
\eta_{AB} h^D_D)
\label{GAB}
\eeq
while $G_{\mu\nu}^{(4)}$ remains in its general form
\beq
\label{einsteinstr}
   G_{\mu\nu}^{(4)} = \frac{1}{2}[
\partial_\mu \partial_\lambda h^\lambda_\nu
+ \partial_\nu \partial_\lambda h^\lambda_\mu
- \Box^{(4)}h_{\mu\nu} - \partial_\mu \partial_\nu h^\lambda_\lambda
-\eta_{\mu\nu}(\partial_\lambda \partial_\rho h^{\lambda \rho}
- \Box^{(4)} h^\lambda_\lambda)]
\eeq
Hereafter 
%$\Box^{(4+N)} = \partial_C \partial^C$ 
%and
$\Box^{(4)} = \partial_\mu \partial^\mu$. 

\subsection{(Quasi-)localized states: tachyon and massive graviton}

Let us consider the 
sourceless field equation, i.e., eq.~(\ref{maineqn})
with $T_{AB} = 0$, to see whether there exist modes which are
(quasi-)localized near the brane.
The $(ab)$ and $(a\mu)$-components of this equation
in the gauge (\ref{gauge}) read
\[
     - \frac{1}{2} {\cal F}(\Box^{(4+N)}) \Box^{(4+N)} (h_{ab} - \frac{1}{2}
\eta_{ab} h^D_D) = 0
\]
\[
 - \frac{1}{2} {\cal F}(\Box^{(4+N)})\Box^{(4+N)} h_{a\mu} = 0 
\]
These are the equations of the bulk theory for the corresponding
combinations of metrics, and they do not have localized
solutions.
 Hence,
\beq
  h_{ab} = \frac{1}{2} \eta_{ab} h_{C}^C
\label{tmp21}
\eeq
and 
\beq
h_{a \mu} = 0
\label{hamu}
\eeq
 After taking trace of eq.~(\ref{tmp21}), 
one expresses the $(ab)$-components of the metric
in terms of the trace of the four-components,
\beq
  h_{ab} = - \frac{1}{N-2} \eta_{ab} h^\mu_\mu
\label{hab}
\eeq
(at this point we specialize to $N>2$).
%One has also
%\beq
%   h_C^C = - \frac{2}{N-2} h^\mu_\mu
%\label{CC}
%\eeq
Then the gauge condition (\ref{gauge}) with $B=\mu$, 
together with eq.~(\ref{hamu})
give
\[
   \partial_\mu h^\mu_\nu = - \frac{1}{N-2} \partial_\nu h^\lambda_\lambda
\]
%where we made use of eq.~(\ref{hamu}).
Making use of the above relations, one
obtains for the remaining, $(\mu\nu)$-components of the field equations
\begin{eqnarray}
&-&\frac{1}{2} {\cal F}(\Box^{(4+N)}) \Box^{(4+N)}
\left(h_{\mu\nu} + \frac{1}{N-2} \eta_{\mu \nu} h^\lambda_\lambda\right)
\nonumber \\
&+& \frac{M_{Pl}}{2} f^2(y) \cdot \int~dy'~f^2 
\left[-\frac{N}{N-2} \partial_\mu \partial_\nu h^\lambda_\lambda
- \Box^{(4)} h_{\mu \nu} + \frac{N-1}{N-2} \eta_{\mu \nu} 
\Box^{(4)} h_\lambda^\lambda\right] = 0
\label{munu1eq}
\end{eqnarray}
Trace of this equation gives
\beq
- {\cal F}(\Box^{(4+N)}) \Box^{(4+N)}
h_{\mu}^\mu 
+ \tilde{M}_{Pl}^2 f^2(y) \int~dy'~f^2 \cdot
 \Box^{(4)} h_{\mu}^\mu
= 0
\label{mumueq}
\eeq 
where
\beq
     \tilde{M}_{Pl}^2 = \frac{2(N-1)}{N+2} M_{Pl}^2
\label{tildeM}
\eeq
The latter is a scalar equation, and we are interested in its solution
localized near the brane. 
This solution is expressed in terms of functions $D_f(p,y)$ and
$D_{ff}(p)$ introduced in eqs.~(\ref{Du}) and (\ref{Duv}).
In the mixed representation the solution is
\beq
  h_\mu^\mu (y) = c \cdot D_f (p^2=m_*^2, y)
\label{wfmumu}
\eeq
where $c$ is a normalization constant and
the mass is determined by the ``eigenvalue equation''
\beq
   m_*^2 = \frac{1}{\tilde{M}_{Pl}^2  D_{ff} (m_*^2)}
\label{eigen}
\eeq
Let us see that the mass squared, $m_*^2$, is, in fact, negative
and real,
\beq
\label{47}
    m_{tachyon}^2 \equiv m_*^2 < 0
\eeq
\beq
\label{48}
   \mbox{Im} (m_*^2) =0
\eeq
so the mode we consider is a tachyon localized near the 
brane.
We first note, that
\[
|m_*| \sim \frac{M_*^2}{M_{Pl}} \sim r_c^{-1}
\]
which is small compared to $M_*$. Now, one has
\beq
  D_{ff} (p^2) 
= - \int~d^N p_y \frac{|f^2(p_y)|^2}{(-P^2) {\cal F}(-P^2)}
\label{Dff1}
\eeq
where $P^2 = p^2 - p_y^2$, as before. 
Since one assumes that the form-factor
${\cal F}$
rapidly grows at large negative $P^2$ (the propagator
$D_*$ rapidly decays), this integral is convergent
in the ulraviolet, and the integrand does not have singularities at
 $p^2 <0$ (a zero of ${\cal F} (-P^2)$ at negative $P^2$ 
would imply that there is a tachyon in bulk theory). 
For $N > 2$ the integral is infrared-convergent even at $p^2=0$, 
since for  $|P^2| \ll M_*^2$ the form-factor ${\cal F}$ is
constant. For small $p^2 <0$ the integral here is a real positive
constant,
which is 
of order $M_*^{-4}$ on dimensional grounds, so 
$D_{ff}(p^2 = -|m_*^2|)$ is a negative constant at small $|m_*|$. 
One concludes
that, as long as scales lower than $M_*$ are concerned, 
there exists a single solution to eq.~(\ref{eigen})
which indeed obeys (\ref{47}), (\ref{48}).

Finally, we have to show that the wave function
(\ref{wfmumu}) decays as $|y| \to \infty$.
One writes
\[
  D_f(p,y) = \int~d^N p_y~
\frac{\mbox{e}^{-i p_y\cdot y} f^2(p_y)}{P^2 {\cal F}(-P^2)}
\]
Large $|y|$ correspond to small $p_y$, so at large
$|y|$ one has
\beq
  D_f (p^2 = m_*^2, y)
\propto - \int~d^N p_y~
\frac{\mbox{e}^{-i p_y\cdot y} f^2(p_y)}{p_y^2 + |m_*^2|}
\label{wfint}
\eeq
Recalling that $f^2(p_y =0) = \int~d^Ny f^2(y) =1$,
one obtains that the wave function (\ref{wfint}) has the shape
of $N$-dimensional Yukawa potential with (small) mass $|m_*|$.
Hence the wave function indeed decays as $|y| \to \infty$.

To obtain the complete tensor structure of the tachyon mode,
one plugs the solution for the trace, eq.~(\ref{wfmumu}),
back into eq.~(\ref{munu1eq}), and obtains, in 
mixed representation,
\begin{eqnarray}
  &-&{\cal F}(\Box^{(4+N)}) \Box^{(4+N)} h_{\mu \nu}
  + M_{Pl}^2 f^2(y) p^2 \int~dy' f^2(y') h_{\mu \nu}(p,y') 
\nonumber \\  
&=& - c \cdot M_{Pl}^2 D_{ff} p^2
\left[\frac{N}{N-2} \frac{p_{\mu} p_\nu}{p^2} -
\frac{N(N-1)}{(N-2)(N+2)} \eta_{\mu \nu} \right]
D_f(p,y)
\end{eqnarray}
where $p^2 = m_*^2$ for the mode we discuss.
This inhomogeneous equation is readily solved.
The $ab$-components are found from eq.~(\ref{hab}).
In this way one finds the complete expression
for the (unnormalized) tachyon mode
\begin{eqnarray}
   h_{\mu \nu}^{(m_*)} &=& \frac{1}{3}
\left(\eta_{\mu\nu} - \frac{N+2}{N-1} \frac{p_\mu p_\nu}{p^2} 
\right) D_f (y)
\nonumber \\
   h_{\mu a}^{(m_*)} &=& 0
\nonumber \\
h_{ab}^{(m_*)} &=& - \frac{1}{N-1} \eta_{ab} D_f (y)
\label{totwf}
\end{eqnarray}
where
$p^2 = m_*^2 < 0$
and
  $ D_f (y) = D_f (p^2 = m_*^2; y)$.

For completeness, let us consider 
(quasi-)localized traceless 
modes, for which $h_\mu^\mu=0$.
For these modes, one obtains from eq.~(\ref{munu1eq}) the following 
equation,
\[
    -{\cal F}(\Box^{(4+N)}) \Box^{(4+N)} h_{\mu \nu} +
   M_{Pl}^2 f^2(y) p^2 \int~dy' f^2(y') h_{\mu \nu}(p,y') = 0
\]
The solution to this equation is again of the form
\[
  h_{\mu \nu} (y) = c_{\mu \nu} \cdot D_f(p^2 = m^2, y)
\]
where $c_{\mu \nu}$ are independent of $y$, and the mass now obeys
\[
   m^2 = - \frac{1}{M_{Pl}^2 D_{ff}(m^2)}
\]
We are interested in solutions with $|m| \ll M_*$, which are
relevant at low energies. 
According to eq.~(\ref{Dff1}), for such a solution
the real part of $m^2$ is positive,
and is of order $r_c^{-2}$. Now, for small positive $p^2$, the 
function $D_{ff}(p^2)$ has even smaller imaginary part,
which may be estimated as follows. The integrand in eq.~(\ref{Dff1})
is a smooth positive function at $p_y^2 \gg p^2$, so this region does
not contribute to the imaginary part. The imaginary part comes from
the infrared region, and is proportional to
\[
\int_0^{\epsilon}~ \frac{p_y^{N-1}~ dp_y}{p_y^2 - p^2 - i0}
\]
The imaginary part of the latter integral is proportional to
\beq
   - i p^{N-2}
\label{impart}
\eeq
So, for $N>2$ the quasi-localized graviton has a small mass
$m_{graviton} \equiv m$, where
\[
   \mbox{Re}(m) = \frac{1}{M_{Pl}\sqrt{|D_{ff}(0)|}}
\sim \frac{M_*^2}{M_{Pl}} \sim r_c^{-1}
\]
and even smaller width,
\[
    \frac{\Gamma_{graviton}}{m} \sim \frac{m^{N-2}}{M_*^{N-2}}
\]
We conclude that in this model, 
there is a massive four-dimensional graviton with a tiny width.
The violation of Newton's law at distances of order $r_c$ is due to
the graviton {\it mass}, not width, in clear contrast to the
five-dimensional case~\cite{DGP}.

\subsection{Propagators at low energies: the tachyon is a ghost}

One way to see that the tachyon is in fact a ghost, is to calculate
the full propagator $D_{AB,CD} (p;y,y')$ near $p^2 = m_*^2$, i.e., 
extract its pole term. This is done in Appendix A. The outcome is
\beq
  D_{AB,CD}^{(pole)}(p;y,y')
  = - \frac{3}{M_{Pl}^2 [D_{ff}(m_*^2)]^2} \cdot
\frac{h_{AB}^{(m_*)} (y) h_{CD}^{(m_*)} (y)}{p^2 - m_*^2}
\label{pole}
\eeq
where $h_{AB}^{(m_*)}(y)$ is the (unnormalized) tachyon wave function
(\ref{totwf}). The overall negative sign here means that
the tachyon is indeed a ghost. 

The  structure of the pole term (\ref{pole})
is precisely what one expects
for the contribution of a mode localized near the brane. From
eq.~(\ref{pole}) one deduces also that the properly normalized
tachyon-ghost mode is
\[
  h_{AB}^{normalized}(y) = \frac{\sqrt{3}}{M_{Pl} |D_{ff}|} 
h_{AB}^{(m_*)}
\]
One observes from the latter formula and eq.~(\ref{totwf}) that
the tachyonic ghost couples to matter on the brane at gravitational
strength. 

%\section{Propagator from brane to brane: how the right tensor
%structure emerges at intermediate scales}

It is perhaps more 
instructive to study the propagator with both end-points
on the brane. More precisely, let us consider the source on the brane
with the only
non-vanishing components $T_{\mu \nu}$, which is distributed 
in the transverse directions with the same\footnote{The analysis 
of the general case of 
a source with spread function $g^2$ different from $f^2$ proceeds 
along the lines of Section 2. This analysis is straightforward
but not illuminating.}
 spread function $f^2(y)$ as
in eq.~(\ref{maineqn}),
\beq
\label{mnsource}
     T_{\mu \nu}(x,y) = \theta_{\mu \nu}(x) f^2(y)\;,
\eeq
where $\theta_{\mu\nu}(x)$ is conserved in the four-dimensional sense.
The point is to calculate $(\mu \nu)$-components
of the metric due to this source.
This is done in Appendix B, with the result, in mixed representation,
\begin{eqnarray}
h_{\mu \nu}(p,y)
&=& \frac{2}{M_{Pl}^2}~\frac{D_f(p,y)}{D_{ff}(p)}~
\left[ \frac{1}{p^2 - m^2(p)} \left(\theta_{\mu \nu} - \frac{1}{3}
\eta_{\mu\nu} \theta_\lambda^\lambda \right)
-\frac{1}{6}~ \frac{1}{p^2 - m_*^2 (p)}~ 
\eta_{\mu\nu} \theta_\lambda^\lambda \right]
\nonumber \\
&+& \mbox{longitudinal~~part}
\label{prop4a}
\end{eqnarray}
% The interaction of these sources is described by
%the propagator (in four-dimensional momentum representation)
%$D^{(4)}_{\mu\nu,\lambda\rho}(p)$ which is a convolution of the
%full propagator with $f^2(y)$ and $f^2 (y')$. The corresponding 
%calculation is sketched in Appendix B, and here we give the result,
%\begin{eqnarray}
%   D^{(4)}_{\mu \nu, \lambda \rho}
%&=& \frac{2}{M_{Pl}^2}
%\left[ \frac{1}{p^2 - m^2 (p)}~\left(\frac{1}{2}
%(\eta_{\mu\lambda} \eta_{\nu \rho} + 
%\eta_{\mu \rho} \eta_{\nu \lambda}) - 
%\frac{1}{3} \eta_{\mu \nu} \eta_{\lambda \rho} \right)
%- 
%\frac{1}{6}~\frac{1}{p^2 - m_*^2 (p)}~\eta_{\mu \nu} \eta_{\lambda \rho}
%\right]
%\nonumber \\
%&+& \mbox{longitudinal~~part}
%\label{prop4}
%\end{eqnarray}
where
\beq
   m^2 (p) = - \frac{1}{M_{Pl}^2 D_{ff}(p)}
\label{mp}
\eeq
\beq
   m_*^2 (p) = \frac{1}{\tilde{M}_{Pl}^2 D_{ff}(p)}
\label{m*p}
\eeq
and the longitudinal part is proportional to $p_\mu p_\nu$
and vanishes when  contracted with
conserved stress-energy (the overall factor $2$ in (\ref{prop4a})
is due to
our definition of $M_{Pl}$, see eq.~(\ref{maineqn})).
Now, the interaction between two sources of the form (\ref{mnsource})
may be written in terms of the effective four-dimensional propagator 
$D^{(4)}_{\mu\nu,\lambda\rho}(p)$, so that one has
\[
   \theta^{\prime}_{\mu\nu} (p) 
D^{(4)}_{\mu\nu,\lambda\rho}(p) \theta_{\lambda \rho}(p)
= \theta^{\prime}_{\mu \nu} (p) \int~d^Ny~f^2(y) h_{\mu \nu}(y,p)\;,
\]
where $h_{\mu\nu}$ is given by eq.~(\ref{prop4a}).
Hence, the effective brane-to-brane propagator is
\begin{eqnarray}
   D^{(4)}_{\mu \nu, \lambda \rho}
&=& \frac{2}{M_{Pl}^2}
\left[ \frac{1}{p^2 - m^2 (p)}~\left(\frac{1}{2}
(\eta_{\mu\lambda} \eta_{\nu \rho} + 
\eta_{\mu \rho} \eta_{\nu \lambda}) - 
\frac{1}{3} \eta_{\mu \nu} \eta_{\lambda \rho} \right)
- 
\frac{1}{6}~\frac{1}{p^2 - m_*^2 (p)}~\eta_{\mu \nu} \eta_{\lambda \rho}
\right]
\nonumber \\
&+& \mbox{longitudinal~~part}
\label{prop4}
\end{eqnarray} 
At low energies the  ``masses'' $m^2(p)$ and $m_*^2(p)$
are constants
(up to tiny $p$-dependent imaginary part, see eq.~(\ref{impart})).
Thus, at low energies the propagator (\ref{prop4}) 
corresponds to a massive graviton of mass $m$ (note that the
Van~Dam--Veltman--Zakharov property indeed holds) and tachyonic ghost
with negative  $m_*^2$. This ghost
cancels the contribution of the extra graviton polarization
at
intermediate momenta $M_* \gg |p| \gg (m,m_*) \sim r_c^{-1}$, so 
that at these scales the brane-to-brane propagator has the same
structure as in General Relativity,
\[
D^{(4)}_{\mu \nu, \lambda \rho}
= \frac{1}{M_{Pl}^2}~\frac{1}{p^2}
(\eta_{\mu\lambda} \eta_{\nu \rho} + 
\eta_{\mu \rho} \eta_{\nu \lambda} - 
\eta_{\mu \nu} \eta_{\lambda \rho})
+ \mbox{longitudinal~~part}
\]
This is precisely the same situation as 
in the model of Ref.~\cite{GRS}: the reason of why the correct tensor
structure emerges in the linearized theory at intermediate distances
is the existence of a ghost field.

\subsection{N=2}

The case $N=2$ is somewhat special. Let us first consider the 
tachyon mode.
Equation (\ref{tmp21}) implies now
\beq
   h_\mu^\mu = 0
\label{tmp111}
\eeq
while $h_{a}^a$ is arbitrary at this point. The four-dimensional
trace of the sourceless equation (\ref{maineqn}) gives then
\[
- {\cal F}(\Box^{(4+N)}) \Box^{(4+N)}
h_{a}^a 
+ \tilde{M}_{Pl}^2 f^2(y) \int~dy'~f^2 \cdot
 \Box^{(4)} h_{a}^a
= 0
\]
This equation has the same structure as eq.~(\ref{mumueq}),
so there again exists a tachyon. 
At $N=2$, it is the extra-dimensional metric $h_{ab}$ and traceless
part of $h_{\mu \nu}$ that do not vanish in the tachyon mode
(in the gauge (\ref{gauge})). 

Another point is that the integral (\ref{Dff1}) is logarithmic
at $N=2$, so the estimate for the graviton and tachyon 
masses is now
\[
     \mbox{Re}(m^2) \sim |m_*^2| 
\sim \frac{M_*^4}{M_{Pl}^2} \log \frac{M_{Pl}}{M_*}
\]
The imaginary part of the graviton mass is suppressed relative to its real
part by logarithm only, 
\[
     \frac{\Gamma_{grav}}{|m|} \sim \frac{1}{\log \frac{M_{Pl}}{M_*}}
\]
Yet the graviton width is smaller than its mass.

The tachyon is a ghost at $N=2$ as well. A simple way
to see this is to redo the calculation leading to the 
brane-to-brane propagator. One finds that the expression
(\ref{prop4}) remains valid at $N=2$, the property (\ref{tmp111})
being ensured by the appropriate structure of the
longitudinal terms. The negative sign of the last term on
the right hand side of eq.~(\ref{prop4}) tells that the tachyon
is indeed a ghost.

So, in spite of peculiarities, the conclusion for $N=2$ is the same as
for $N>2$: the model has a tachyonic ghost.

\section{Generalized model}
In this section we drop the assumption that the
tensor structure of the linearized bulk equations 
coincides with
that in the linearized Einstein theory 
and consider the
most general tensor structure compatible with the $(4+N)$-dimensional
general covariance.
%transformations of the metric,
%\beq
%\label{trans}
%h_{AB} \to h_{AB}+\d_A \xi_B+\d_B \xi_A\;.
%\eeq
The linearized field equation in the bulk
theory has the following general form,
\beq
{\cal D}_{ABCD}h^{CD}=0
\eeq
with some linear  operator ${\cal D}_{ABCD}$. The symmetry of
this operator under  $A\leftrightarrow B$,
$C\leftrightarrow D$ and $(AB)\leftrightarrow (CD)$ implies the
following structure  of ${\cal D}_{ABCD}$ 
\begin{eqnarray}
{\cal D}_{ABCD}=a \d_A\d_B\d_C\d_D+
b\l \d_A\d_B\eta_{CD}+\eta_{AB}\d_C\d_D\r
\nonumber\\
+c\l\d_A\eta_{BC}\d_D+
\d_B\eta_{AC}\d_D+\d_A\eta_{BD}\d_C+\d_B\eta_{AD}\d_C\r 
\nonumber \\+
d\eta_{AB}\eta_{CD}+
e\l\eta_{AC}\eta_{BD}+\eta_{AD}\eta_{BC}\r\;,\nonumber
\end{eqnarray}
where $a,b,c,d,e$ are yet arbitrary functions of $\Box^{(N+4)}$.
Now, gauge invariance implies
\[
\d_A{\cal D}^A_{BCD}=0\;,
\]
This
leaves only two possible tensor structures which may appear in 
${\cal D}_{ABCD}$, namely the usual Einstein 
structure 
and the product of two projectors, ${\cal D}_{ABCD} \propto P_{AB}P_{CD}$,
where
\[
P_{AB}=\Box^{(4+N)}\eta_{AB}-\d_A\d_B
\]
%\[
%P_{AB}=(\Box^{(4+N)}\eta_{AB}-\d_A\d_B)(\Box^{(4+N)}\eta_{CD}-\d_C\d_D)
%h_{CD}\;.
%\]
In the harmonic gauge (\ref{gauge}) one has
\[
P_{AB} P_{CD}h^{CD} \equiv \Pi_{AB}
={1\over 2}(\Box^{(4+N)}\eta_{AB}-\d_A\d_B)\Box^{(4+N)} h^C_C\;.
\]
Consequently, the generalization of
eq. (\ref{maineqn}) is
\beq
   {\cal F}(\Box^{(4+N)}) G_{AB} (x,y) +{\cal G}(\Box^{(4+N)}) 
\Pi_{AB} (x,y)+ M_{Pl}^2 f^2(y)
\int~dy'~f^2(y') G^{(4)}_{AB} (x, y') = T_{AB}(x,y)\;,
\label{maineqn1}
\eeq
where $G_{AB}$ is given by eq.~(\ref{GAB}), and
a new form-factor ${\cal G}$ is assumed to have 
the same ultraviolet properties as the form-factor ${\cal F}$.

We will not repeat all the steps of the 
analysis of Section 3. To see the existence of the localized
ghost and find its wave function in this general setup,
it suffices to study the 
structure of the brane-to-brane propagator.
We again consider a source of the form (\ref{mnsource})
and evaluate 
%\beq
%\label{mnsource}
%     T_{\mu \nu}(x,y) = \theta_{\mu \nu}(x) f^2(y)\;.
%\eeq
the $(\mu\nu)$-components of the metric induced by this source. 
The result is (see Appendix C for calculational details)
\begin{eqnarray}
\label{genhmn}
h_{\mu \nu}(p,y)
&=& \frac{2}{M_{Pl}^2}\left[~\frac{D_f(p,y)}{D_{ff}(p)}~
\frac{1}{p^2 - m^2(p)} \left(\theta_{\mu \nu} - \frac{1}{3}
\eta_{\mu\nu} \theta_\lambda^\lambda \right)
-{\tilde{D}_f(p,y)\over \tilde{D}_{ff}}\frac{1}{6}~ \frac{1}{p^2 - m_*^2 (p)}~ 
\eta_{\mu\nu} \theta_\lambda^\lambda \right]
\nonumber \\
&+& \mbox{longitudinal~~part}
\end{eqnarray}
where  $\tilde{D}_f$ and $\tilde{D}_{ff}$ are defined in a similar way as
$D_f$ and $D_{ff}$ (see eqs. (\ref{Du}), (\ref{Duv})) 
but with the a new 
function $\tilde{D}_*(p;y,y')$
substituted for  $D_*(p;y,y')$. The  function $\tilde{D}_*(p;y,y')$
is a solution of the
following equation
\[
 - {\cal O}(\Box^{(4+N)})\cdot
{\cal F}(\Box^{(4+N)}) \cdot \Box^{(4+N)} \tilde{D}_* (p;y,y') = \delta (y-y')\;,
\]
where the operator ${\cal O}(\Box^{(4+N)})$ is 
\[
{\cal O}(\Box^{(4+N)})={N-1+N{\cal H}(\Box^{(4+N)})\cdot
\Box^{(4+N)}\over
N+2-(N+3){\cal H}(\Box^{(4+N)})\cdot
\Box^{(4+N)}}
\]
with
\[
{\cal H}(\Box^{(4+N)})={2{\cal G}(\Box^{(4+N)})\over {\cal
F}(\Box^{(4+N)})}
\;. 
\]
The ``mass'' $m^2(p)$ entering eq.~(\ref{genhmn})
is the same as in Section 3, while $m_*^2 (p)$ is now 
\[
    m_*^2(p) = \frac{1}{2{M}_{Pl}^2  \tilde{D}_{ff} (p^2)}
\]
Thus, the brane-to-brane propagator still has the form (\ref{prop4}).
The second term in eq.~(\ref{prop4}) again has negative  sign,
so the model again has ghost field,
but now the mass of the ghost is
a
solution to the following eigenvalue equation
\beq
   m_*^2 = \frac{1}{2{M}_{Pl}^2  \tilde{D}_{ff} (\tilde{m_*^2)}}\;.
\label{eigentilde}
\eeq
The difference with the case ${\cal G}=0$ studied in Section 3 is
that the ghost field in principle need not be a tachyon in the general case
and that the wave functions of the ghost and graviton  have
different profiles in the transverse directions. The 
graviton wave function is again $D_f(p;y)$, 
while the ghost wave function is $\tilde{D}_f(p;y)$.

\section*{Acknowledgements}
The authors are indebted to I.~Antoniadis, A.~Barvinsky, R.~Rattazzi, 
S.~Sibiryakov, M.~Shaposhnikov, M.~Shifman, P.~Tinyakov
and F.~Viniegra for helpful
discussions. We are grateful to 
G.~Gabadadze for stimulating
correspondence. This work has been supported in part by
Russian Foundation for Basic Research, grant 02-02-17398,
Swiss Science Foundation grant 7SUPJ062239, and the CRDF grant 
RP1-2364-MO-02.
The work of S.D. has been supported in part 
by the INTAS grant YS 2001-2/128.

\section*{Appendix A}

Here we calculate the tachyon
pole term in the full propagator of the model of Section 3.
We do this by solving eq.~(\ref{maineqn}) with conserved
right hand side,
\beq
  \partial_A T^A_B = 0
\label{apcons}
\eeq 
Otherwise $T_{AB}$ is arbitrary. We still use the gauge conditions
(\ref{gauge}). 

We begin with $(ab)$-components of eq.~(\ref{maineqn}) which read
\beq
     - \frac{1}{2} {\cal F} \Box^{(4+N)} (h_{ab} - \frac{1}{2}
\eta_{ab} h^D_D) = T_{ab}
\label{apab}
\eeq
We decompose $h_{ab} $ in the following way,
\beq
  h_{ab} = h_{ab}^T + \frac{1}{N} \eta_{ab} h^c_c
\eeq
where
\beq
  h_{ab}^T =h_{ab} - \frac{1}{N} \eta_{ab} h^c_c
\eeq
is the traceless part.
The traceless part obeys
\beq
   -\frac{1}{2} {\cal F}(\Box^{(4+N)}) \Box^{(4+N)} h_{ab}^{T}
= T_{ab} - \frac{1}{N} \eta_{ab} T^c_c
\label{hTeq}
\eeq
while the trace of eq.~(\ref{apab}) gives
\beq
  h_c^c = - \frac{N}{N-2} h^\mu_\mu + b
\label{tracedef}
\eeq
where $b$ obeys
\beq
 -{\cal F}(\Box^{(4+N)}) \Box^{(4+N)} b
= -\frac{4}{N-2} T_{c}^c  
\label{beq}
\eeq
and hence is equal to
\beq
  b(X) =-\frac{4}{N-2} \int~d^{4+N}X' ~ D_*(X-X') 
T_{c}^c (X')   
\eeq
We also have for the overall trace
  \beq
   h_C^C = - \frac{2}{N-2} h^\mu_\mu +b
\eeq
Let us now consider $(a\mu)$-components of eq.~(\ref{maineqn}).
They read
\beq
  -\frac{1}{2} {\cal F}(\Box^{(4+N)}) \Box^{(4+N)} h_{a \mu}
= T_{a \mu}
\label{hamueq}
\eeq
Hence,
\beq
  h_{a\mu} (X) = 2 \int~d^{4+N} X'~ D_*(X-X') 
  T_{a\mu} (X')   
\eeq
Finally, let us study $(\mu\nu)$-components of eq.~(\ref{maineqn}).
We need the expression for $\partial_\lambda h^{\lambda \rho}$ 
which enters
$G^{(4)}_{\mu \nu}$. This expression is obtained by making use of the
gauge condition
\beq
  \partial_A h^A_\mu 
\equiv \partial_a h^a_\mu + \partial_\lambda h^\lambda_\mu
= \frac{1}{2} \partial_\mu h_C^C
\label{gaugecond}
\eeq
Now, because of the conservation property (\ref{apcons}),
one has
\beq
  \partial_a h^a_\mu  = 2 \int~d^{4+N} X'~ D_*(X-X') 
  \partial_a T_{\mu}^a (X') =
  - 2 \int~d^{4+N} X'~ D_*(X-X') 
  \partial_\lambda T_{\mu}^\lambda (X')
\eeq
Hence, $(\mu\nu)$-components of eq.~(\ref{maineqn})
may be written in terms of $h_{\mu\nu}$, the trace
$h_\lambda^\lambda$ and components $T_\lambda^\rho$ and $T_a^a$ of the
stress-energy tensor. After some algebra one obtains
\begin{eqnarray}
&-& {\cal F}(\Box^{(4+N)}) \Box^{(4+N)}
\left(h_{\mu\nu} + \frac{1}{N-2} \eta_{\mu \nu} h^\lambda_\lambda\right)
\nonumber \\
&+& M_{Pl}^2 f^2(y) \cdot \int~dy'~f^2 (y') 
\left[-\frac{N}{N-2} \partial_\mu \partial_\nu h^\lambda_\lambda
- \Box^{(4)} h_{\mu \nu} + \frac{N-1}{N-2} \eta_{\mu \nu} 
\Box^{(4)} h_\lambda^\lambda\right] 
\nonumber \\
&=& 2 T_{\mu\nu} - \frac{2}{N-2}\eta_{\mu\nu} T^a_a
\nonumber \\
&-& 2 M_{Pl}^2 f^2(y) \int~d^Ny'~D_f(y')
\left[ \partial_\lambda \partial_\mu T^\lambda_\nu
+ \partial_\lambda \partial_\nu T^\lambda_\mu 
-\frac{2}{N-2} \partial_\mu \partial_\nu T^a_a \right.
\nonumber \\
 &-& \left.  \eta_{\mu \nu} 
\left(\partial_\lambda \partial_\rho T^{\lambda \rho}
-\frac{1}{N-2} \Box^{(4)} T_a^a \right)\right]
\label{apmunueq}
\end{eqnarray}
Trace of this equation gives
\begin{eqnarray}
&-& {\cal F}(\Box^{(4+N)}) \Box^{(4+N)}
h_{\mu}^\mu 
+ \tilde{M}_{Pl}^2 f^2(y) \int~dy'~f^2 \cdot
 \Box^{(4)} h_{\mu}^\mu
\nonumber \\
&=&  \frac{2(N-2)}{N+2}\left[ T_\mu^\mu -\frac{4}{N-2}T_a^a
\right.
\nonumber \\
&+& \left. 2 M_{Pl}^2 f^2(y) \int~d^Ny'~D_f(y') \cdot 
\left(\partial_\lambda \partial_\rho T^{\lambda \rho}
-\frac{1}{N-2} \Box^{(4)} T^a_a \right) \right]
\label{apmumueq}
\end{eqnarray}
The solution to the latter equation is conveniently written in mixed
representation, 
\begin{eqnarray}
h^\mu_\mu (p,y) &=&\frac{2(N-2)}{N+2} \int~d^Ny'~D_*(p;y,y')
\left(T_\mu^\mu (p,y')- \frac{4}{N-2} T_a^a (p,y') \right)
\nonumber \\
&+& \frac{4(N-2)}{N+2} 
\frac{p^2 M_{Pl}^2 D_f(y)}{1 - \tilde{M}_{Pl}^2 p^2 D_{ff}(p)}
\nonumber \\
&\times&
\int~d^Ny'~D_f(p,y') \left( \frac{N-1}{N+2} T_\mu^\mu
- \frac{p_\mu p_\nu}{p^2} T^{\mu \nu} -\frac{3}{N+2} T^a_a \right)
\label{aptraceh}
\end{eqnarray}
where we made use of the relation (\ref{tildeM}). This expression is
still exact.
Clearly, it has a pole at $p^2 = m_*^2$
(in the second term in the right hand side of 
eq.~(\ref{aptraceh})).

To find the tachyon
pole term in $h_{\mu\nu}$, one plugs the solution for
$h_\lambda^\lambda$ back into eq.~(\ref{apmunueq}). 
In terms of eq.~(\ref{apmunueq}), the tachyon pole
in $h_{\mu\nu}$ comes entirely from the pole part in
$h_\lambda^\lambda$.
One makes use of eq.~(\ref{apmumueq}) and writes eq.~(\ref{apmunueq})
in the following form
\begin{eqnarray}
 &-&{\cal F}(\Box^{(4+N)}) \Box^{(4+N)} h_{\mu \nu}
  + M_{Pl}^2 f^2(y) p^2 \int~dy' f^2(y') h_{\mu \nu}(p,y') 
\nonumber \\  
&=&
M_{Pl}^2 p^2 f^2(y) \int~d^Ny'~f^2 \left(
\frac{N(N-1)}{(N-2)(N+2)} \eta_{\mu \nu} 
-\frac{N}{N-2} \frac{p_{\mu} p_\nu}{p^2} \right)
h^\lambda_\lambda + \dots
\end{eqnarray}
where dots denote terms that do not contain a pole at $p^2 = m_*^2$.
This equation is straightforwardly solved, and after some algebra 
one obtains that the tachyon
pole part of $h_{\mu\nu}$ is 
\begin{eqnarray}
h_{\mu \nu}^{(pole)} (p,y)
&=&
- \frac{1}{3}~\frac{1}{p^2 - m_*^2}~
\frac{D_f(p,y)}{M_{Pl}^2D_{ff}^2}
\left(\eta_{\mu\nu} - \frac{N+2}{N-1} \frac{p_\mu p_\nu}{m_*^2}
\right)
\nonumber \\
&\times& \int~d^Ny' ~D_f(p,y')\left[\left(\eta_{\lambda\rho} - 
\frac{N+2}{N-1} \frac{p_\lambda p_\rho}{m_*^2} 
\right) T^{\lambda \rho}(p,y')
-\frac{3}{N-1} T^a_a (p,y') \right]
\label{apo1}
\end{eqnarray}
It remains to find the tachyon
pole parts of other metric components.
According to eqs.~(\ref{hTeq}), (\ref{beq}) and (\ref{hamueq}),
the traceless part $h_{ab}^T$, the term $b$ and the metric
components $h_{a\mu}$ do not have poles at $p^2 = m_*^2$.
The pole term in $h^a_a$ is determined by the pole term in $h_\mu^\mu$
through eq.~(\ref{tracedef}). Thus,
 one finds 
\begin{eqnarray}
 h_{ab}^{(pole)} (p,y) &=& - 
\frac{1}{N-2} \eta_{ab} h^{(pole)\mu}_{~~~~\mu} (p,y)
\nonumber \\
&=& 
\frac{1}{N-1}~ \eta_{ab}~\frac{1}{p^2 - m_*^2}~
 \frac{D_f(p,y)}{M_{Pl}^2D_{ff}^2}
\nonumber \\
&\times&
\int~d^Ny' ~D_f(p,y')\left[\left(\eta_{\lambda\rho} - 
\frac{N+2}{N-1} \frac{p_\lambda p_\rho}{m_*^2} 
\right) T^{\lambda \rho}(p,y')
-\frac{3}{N-1} T^a_a (p,y') \right]
\label{apo2}
\end{eqnarray}
We see from eqs.~(\ref{apo1}) and
(\ref{apo2}) that the pole terms in $h_{AB}$ may indeed be written
in the form
\beq
  h_{AB}^{(pole)} (p,y) = \int~d^Ny'~D_{AB,CD}^{(pole)}(p;y,y')
T^{CD}(p,y')
\eeq
where the pole term in the propagator is given by 
eq.~(\ref{pole}).

\section*{Appendix B}

Let us calculate the $(\mu\nu)$-components of the metric due to 
the source of the form (\ref{mnsource})
%\beq
%     T_{\mu \nu}(x,y) = \theta_{\mu \nu}(x) f^2(y)
%\eeq
with conserved $\theta_{\mu\nu}$. For this particular type of source,
the expression (\ref{aptraceh})  simplifies 
considerably,
\beq
  h_\mu^\mu (p,y) =
 -\frac{2(N-2)}{N+2}~\frac{D_f(p,y)}{\tilde{M}_{Pl}^2 D_{ff}(p)}~
\frac{1}{p^2 - m_*^2 (p)} \theta_\mu^\mu (p)
\eeq
where $m_*^2(p)$ is given by eq.~(\ref{m*p}). We plug this 
expression into eq.~(\ref{apmunueq}) and again make use of the
properties of the source to obtain
\begin{eqnarray}
 &-&{\cal F}(\Box^{(4+N)}) \Box^{(4+N)} h_{\mu \nu}
  + M_{Pl}^2 f^2(y) p^2 \int~dy' f^2(y') h_{\mu \nu}(p,y') 
\nonumber \\  
&=&
f^2 (y) \left( 2\theta_{\mu\nu} - 
\frac{2}{N+2} \eta_{\mu \nu} \theta_\lambda^\lambda
-\frac{N}{N+2}~ \frac{p^2}{p^2 - m_*^2 (p)}~
 \eta_{\mu\nu} \theta_\lambda^\lambda \right)
\nonumber \\
&+& \mbox{longitudinal~~part}
\end{eqnarray}
where the longitudinal part is proportional to $p_\mu p_\nu$.
After some algebra, one finds that the solution has indeed the form
(\ref{prop4a}).
%
%writes for
% the solution to the latter equation,
%\begin{eqnarray}
%h_{\mu \nu}(p,y)
%&=& \frac{2}{M_{Pl}^2}~\frac{D_f(p,y)}{D_{ff}(p)}~
%\left[ \frac{1}{p^2 - m^2(p)} \left(\theta_{\mu \nu} - \frac{1}{3}
%\eta_{\mu\nu} \theta_\lambda^\lambda \right)
%-\frac{1}{6}~ \frac{1}{p^2 - m_*^2 (p)}~ 
%\eta_{\mu\nu} \theta_\lambda^\lambda \right]
%\nonumber \\
%&+& \mbox{longitudinal~~part}
%\end{eqnarray}
%This immediately leads to the propagator (\ref{prop4}),
%as the interaction 
%between sources on the
%brane involves the convolution of this metric
%with $f^2(y)$. 

\section*{Appendix C}
Here we sketch the calculations leading to the result
(\ref{genhmn}). The steps are  similar to those in 
Appendices A and B. First, one considers $(ab)$-components of
eq. (\ref{maineqn1}) and finds the following generalization of
eq. (\ref{tracedef}) (recall that we consider a source of the form 
(\ref{mnsource})),
\beq
  h_c^c = - \frac{N+{\cal H}\l \l N-1\r \Box^{(4+N)}+\Box^{(4)}\r }
{N-2+{\cal H}\l \l N-1\r \Box^{(4+N)}+\Box^{(4)}\r} h^\mu_\mu \;.
\label{gtracedef}
\eeq
For the overall trace one has
\beq
\label{ghcc}
   h_C^C = - \frac{2}{N-2+{\cal H}\l \l N-1\r
\Box^{(4+N)}+\Box^{(4)}\r}
 h^\mu_\mu\;.
\eeq
From the $(a\mu)$-components of eq.~(\ref{maineqn1}) one finds
\beq
 h_{a\mu} =  \frac{\cal H}{N-2+{\cal H}\l \l N-1\r
\Box^{(4+N)}+\Box^{(4)}\r}\d_a\d_{\mu}
 h^\nu_\nu\;.
\eeq
By making use of the gauge condition (\ref{gaugecond}) one obtains the
following expression for the longitudinal components of the
four-dimensional part of the metric,
\beq
\label{dnhnm}
\d_{\nu}h^\nu_\mu=-\frac{1-{\cal H}\Box^{(N)}}{N-2+{\cal H}\l \l N-1\r
\Box^{(4+N)}+\Box^{(4)}\r}\d_{\mu}
 h^\nu_\nu\;,
\eeq
where $\Box^{(N)}=\delta^{ab}\d_a\d_b$.
Plugging expressions (\ref{ghcc}), (\ref{dnhnm}) into the 
$(\mu\nu)$-components of  eq.~(\ref{maineqn1}) one arrives at the
following analog of eq.~(\ref{apmunueq}),
\begin{eqnarray}
&-& {\cal F}(\Box^{(4+N)}) \Box^{(4+N)}
\left(h_{\mu\nu} + \frac{1+{\cal H}\Box^{(N+4)}}{N-2+{\cal H}\l \l N-1\r
\Box^{(4+N)}+\Box^{(4)}\r} \eta_{\mu \nu} h^\lambda_\lambda\right)
\nonumber \\
&+& {\cal F}(\Box^{(4+N)}) \Box^{(4+N)}
\frac{{\cal H}}{N-2+{\cal H}\l \l N-1\r
\Box^{(4+N)}+\Box^{(4)}\r}\d_\mu\d_\nu h_\lambda^\lambda
\nonumber \\
&+& M_{Pl}^2 f^2(y) \cdot \int~dy'~f^2 (y') 
\left[-\frac{N+{\cal H}\l \l N+1\r \Box^{(4+N)}-\Box^{(4)}\r }
{N-2+{\cal H}\l \l N-1\r \Box^{(4+N)}+\Box^{(4)}\r} \partial_\mu \partial_\nu
h^\lambda_\lambda\right.
\nonumber\\
&-&\left. \Box^{(4)} h_{\mu \nu} + \frac{N-1
+{\cal H}N\Box^{(4+N)}}
{N-2+{\cal H}\l \l N-1\r \Box^{(4+N)}+\Box^{(4)}\r}\eta_{\mu \nu} 
\Box^{(4)} h_\lambda^\lambda\right] 
%\nonumber \\
= 2 \theta_{\mu\nu}f^2(y) \;.
\label{gapmunueq}
\end{eqnarray}
Trace of this equation gives
\begin{eqnarray}
&-& {\cal F}(\Box^{(4+N)}) \Box^{(4+N)}
\frac{N+2+{\cal H}\l N+3\r \Box^{(4+N)}}{N-2+{\cal H}\l \l N-1\r
\Box^{(4+N)}+\Box^{(4)}\r}
h_{\mu}^\mu 
\nonumber\\
&+&2M_{Pl}^2 f^2(y) \int~dy'~f^2 \cdot
 \Box^{(4)} \frac{N-1+{\cal H} N \Box^{(4+N)}}{N-2+{\cal H}\l \l N-1\r
\Box^{(4+N)}+\Box^{(4)}\r}h_{\mu}^\mu
%\nonumber \\
= 2\theta_\mu^\mu f^2(y)\;.
\label{gapmumueq}
\end{eqnarray}
The solution of this equation is
\beq
h_\mu^\mu={2\theta_\mu^\mu\over 1-2p^2D_{ff}(p)}{N-2+{\cal H}(\l N-1\r
\Box^{(4+N)}+\Box^{(4)}) \over N+2+{\cal H}\l N+3\r \Box^{(4+N)}}\tilde{D}_f\;.
\label{ghmm}
\eeq
Plugging this result back into eq.~(\ref{gapmunueq}) one obtains after some
algebra the desired expression (\ref{genhmn}).

\bibliography{}

\end{document}